\documentclass[%
 reprint,
nofootinbib,
 amsmath,
 amssymb,
 aps,
]{revtex4-1}

\newcommand{\boldnabla}{\mbox{\boldmath$\nabla$}}

\usepackage{color}
\usepackage{graphicx}
\usepackage{dcolumn}
\usepackage{bm}
\usepackage[mathlines]{lineno}
\usepackage[perpage]{footmisc}

\begin{document}

\preprint{APS/123-QED}

\title{Fast chemical reaction in a two-dimensional Navier-Stokes flow: \\
probability distribution in the initial regime}

\author{Farid Ait-Chaalal}
 \email{Corresponding author, farid.aitchaalal@mcgill.ca} 
\author{Michel S. Bourqui}%
\author{Peter Bartello}%
\affiliation{%
McGill University,
Atmospheric and Oceanic Sciences
Room 945, Burnside Hall,
805 Sherbrooke Street West,
Montreal, Quebec H3A 2K6, Canada
}%

\date{\today}

\begin{abstract}
We study an instantaneous bimolecular chemical reaction in a two-dimensional chaotic, incompressible and closed Navier-Stokes flow. 
Areas of well mixed reactants are initially separated by infinite gradients. We focus on the initial regime, characterized by a 
well-defined one-dimensional contact line between the reactants. The amount of reactant consumed is given by the diffusive flux along this line, and 
hence relates directly to its length and to the gradients along it. 
We show both theoretically and numerically that the probability distribution of the
modulus of the gradient of the reactants along this contact line multiplied by $\kappa^{\frac{1}{2}}$ does not depend on the diffusion $\kappa$ and can be inferred, 
after a few turnover times, from the joint distribution of the finite time Lyapunov exponent $\lambda$ and the frequency $\frac{1}{\tau}$. The equivalent time $\tau$ 
measures the stretching time scale of a Lagrangian parcel in the recent past, while $\frac{1}{\lambda}$ measures it on the whole chaotic orbit. 
At smaller times, we predict the shape of this gradient distribution taking
into account the initial random orientation between the contact line and the stretching direction. We also show that the probability distribution of the reactants is
proportional to $\kappa^{\frac{1}{2}}$ and to the product of the ensemble mean contact line length with the ensemble mean of the inverse of the gradient along it. 
Besides contributing to the understanding of fast chemistry in chaotic flows, the present study based on a Lagrangian stretching theory approach provides
results that pave the way to the development of accurate subgrid parametrizations in models with insufficient 
resolution for capturing the length scales relevant to chemical processes, for example in Climate-Chemsitry Models.

\end{abstract}
                             
\maketitle

\section{Introduction}

Chemical reactions in the stratosphere have been shown to be sensitive to the numerical spatial resolution 
when the chemistry is fast compared to advective processes (\cite{Tan1998,Edouard1996}). It was proposed by \cite{Tan1998} that the product 
concentration of the deactivation of polar vortex chlorine by low latitudes nitrogen oxide at the edge of the stratospheric 
Northern hemisphere winter time polar vortex scales like $\kappa^{p(t)}$, with being $\kappa$ the reactant diffusion. Later on, \cite{Wonhas02} argued that $p(t)$ can be
expressed as $1-D(t)/2$, where $D(t)$ is the box counting fractal dimension of the contact line between the reactants.
Here we focus on the initial regime of a instantaneous bimolecular chemical reaction in a two dimensional Navier-Stokes flow characterized by chaotic trajectories 
(this provides an idealized framework for isentropic dynamics in the stratosphere). By definition, the initial regime is characterized by a well-defined one-dimensional contact 
line (i.e. $D=1$). The reactants are initially separated by infinite gradients. 

In a previous work (\cite{Ait12}) dealing with this regime, we have shown that the ensemble mean reactant concentration time derivative scales like $\kappa^\frac{1}{2}$ and  
can be predicted accurately from the Lagrangian stretching properties of the flow. Here we investigate the statistical properties of the chemical production and of the 
reactants concentrations.

In section II, we explain how the study of an infinitely fast chemical reaction $A + B \longrightarrow C$ simplifies into the study of a passive tracer whose concentration field
$\phi$ is defined as the 
difference between the concentrations fields of the two reactants $A$ and $B$ . The rate at which the reactants disappear is the diffusive flux of $\phi$ along the contact line, 
and hence depends on both 
its length and the gradients along it. In section III, we give some theoretical relations between, on one hand, the contact line length and the 
gradients of the reactants along it, and, on the other hand, the Lagrangian stretching properties of the flow. In section IV, we focus on the probability distribution of the 
reactants concentration. Section V describes the Lagrangian stretching properties of a two-dimensional Navier-Stokes flow and section VI presents some numerical simulations to test the 
theoretical results of sections III and IV in the flow introduced in section V.

\section{Finite time Lyapunov exponents and chemical production in a chaotic flow}

We consider the bimolecular chemical reaction $A + B \longrightarrow C$ in stoechiometric quantities. One molecule of $A$ reacts locally with one molecule 
of $B$ to give one molecule of $C$. As a result, the field $\phi=C_A-C_B$, defined as the difference between the reactants' concentration fields 
$C_A$ and $C_B$, is independent of the chemistry. If, in addition, the reaction is instantaneous, 
$A$ and $B$ cannot coexist at the same location. Consequently, the fields $C_A$ and $C_B$, and their spatial average over a closed domain
denoted by an overbar, can be retrieved from $\phi$ as follows (\cite{Ait12,Wonhas02}):
\begin{subequations}\label{E:inf_chim}
 \begin{gather}
  \left\{\begin{matrix}
    C_A(\mathbf{x},t)= \phi(\mathbf{x},t) \mbox{ and } C_B(\mathbf{x},t)=0 & \mbox{     if } \phi(\mathbf{x},t) > 0 \\ 
    C_B(\mathbf{x},t)=-\phi(\mathbf{x},t) \mbox{ and } C_A(\mathbf{x},t)=0 & \mbox{     if } \phi(\mathbf{x},t) < 0 
  \end{matrix}\right. \\
  \overline{C_A} = \overline{C_B}=\frac{\overline{\left |\phi\right |}}{2} \label{E:tot1}
  \end{gather}
\end{subequations}
If $A$ and $B$ are separated by a contact line $\mathcal{L}=\{\mathbf{x}\lvert\phi(\mathbf{x})=0\}$ of dimension one oriented  in a
counterclockwise direction such that
it encloses reactant $A$ (domain $D_A$), the time derivative of the reactants in an incompressible closed flow is: 
\begin{equation}\label{E:difflux}
  \mathcal{A} \frac{\overline{dC_A}}{dt} = \mathcal{A} \frac{\overline{dC_B}}{dt}=\frac{1}{2} \mathcal{A} \frac{d \overline{\left | \phi \right |}}{dt}=
  - \kappa \int_{\mathcal{L}(t)} \nabla \phi \cdot \mathbf{n}dl  \mbox{,}
\end{equation}
where $\mathcal{A}$ is the total area of the domain and $\mathbf{n}$ a unit vector normal to $\mathcal{L}(t)$ pointing outward from $D_A$. 
We call $-\frac{d \overline{\left | \phi \right |}}{dt}$ the chemical speed. Furthermore, on every line element $dl$ of $\mathcal{L}(t)$, the quantities
$\frac{\kappa}{2} \nabla \phi \cdot \mathbf{n}dl$ of $A$ and $B$ are consumed. As a consequence, knowing the length of $\mathcal{L}(t)$ and the probability
distribution of $|\nabla \phi|$ along it, gives a comprehensive statistical description of the chemistry in the domain. 

The equation for the passive tracer $\phi$ is:
\begin{equation}\label{E:tracer}
  \frac{\partial{\phi}}{\partial {t}} + \mathbf{u}\cdot \nabla {\phi} = \kappa \nabla^2  {\phi},                                                    
\end{equation}
where $\mathbf{u}$ is the flow. If the trajectories are chaotic, equation ($\ref{E:difflux}$) allows to link the chemical 
speed to the Lagrangian stretching properties of the trajectories as captured by the finite time Lyapunov exponents (FTLE), defined as the rate of
exponential increase of the distance between the trajectories of two fluid parcels that are initially infinitely close. If 
$\bm{\delta l}(t)$ is the distance between two parcels that start at $\mathbf{x}$ and  $\mathbf{x+\bm{\delta l}_0}$, then the FTLE 
$\lambda(\mathbf{x},t)$ at $\mathbf{x}$ over the time interval $t$ is
\begin{equation}\label{E:ftle_def}
 \lambda(\mathbf{x},t)=\frac{1}{t} \max_{\alpha} \big\{ \ln{\frac{\mathbf{|\bm{\delta l}|}}{\mathbf{|\bm{\delta l}_0|}}} \big\} \mbox{   ,}
\end{equation}
where the maximum is calculated over all the possible orientations $\alpha$ of $\bm{\delta l}_0$. The unit vector with the 
orientation $\psi_{+}(\mathbf{x},t)$ of $\bm{\delta l}_0$ at the maximum defines a ``singular vector'' $\bm{\psi_+}(\mathbf{x},t) \equiv (\cos \psi_{+},\sin \psi_{+})$.
In the inviscid limit, it can be shown, with the conservation of tracer 
concentration for Lagrangian parcels, that $\lambda(\mathbf{x},t)$ is also the rate of exponential increase of a gradient initially aligned with the unit vector
$\bm{\psi_-}(\mathbf{x},t) \equiv (-\sin \psi_{+},\cos \psi_{+})$ perpendicular to $\bm{\psi_+}$.

We can calculate FTLE and singular vectors in an incompressible flow using the velocity gradient tensor $\mathbf{S} \equiv \nabla \mathbf{u}(\mathbf{X},t)$ 
along a trajectory $\mathbf{X}(\mathbf{x},t)$. The distance $\bm{\delta l}$ between two trajectories initially infinitely close is solution of 
$\frac{d\bm{\delta l}}{dt}-\mathbf{S}(t).\bm{\delta l}=0$
and is given by $\bm{\delta l}=\mathbf{M}\bm{\delta l(t=0)}$ 
where the resolvent matrix
$\mathbf{M}$ is solution of $\frac{d\mathbf{M}}{dt}-\mathbf{S}(t).\mathbf{M}=0$. The finite time Lyapunov exponent $\lambda(t)$ is given by the 
$\log$ of the largest eigenvalue of 
$[\mathbf{M}^{T}\mathbf{M}]^{\frac{1}{2t}}$, with $\bm{\psi}_{+}$ the associated eigenvector. 

In ergodic chaotic dynamical systems, it has been shown that the FTLE converge to an infinite time Lyapunov exponent independent of the initial position $\mathbf{x}$, 
while the singular vectors converge to the forward Lyapunov vector $\bm{\Psi_{+}} \equiv (\cos \Psi_{+},\sin \Psi_{+})$ that depend on $\mathbf{x}$ 
(Osedelec theorem, \cite{Os68}). The convergence of the Lyapunov exponent is very slow and algebraic in time 
(\cite{Tang1996}) while the convergence of the Lyapunov vector is much faster, and typically exponential (\cite{Gol87}). A discussion specific to high Reynolds number
two-dimensional Navier-Stokes flows
is available in \cite{Lapeyre02}. Here we will only take 
into account the time dependence of the FTLE, assuming the singular vector field is only a function of space, not of time, equal to the forward Lyapunov vector field. 
This assumption will allow us to link the evolution of a gradient along a trajectory to the Lagrangian straining properties of the flow, 
while taking into account the diffusion (see  ($\ref{E:gradf_general_sol}$) below)

An element $\bm{\delta l_{0}}=\left| \bm{\delta l_{0}} \right|(\cos \alpha,\sin \alpha)$ of the contact line at the initial time is advected at time $t$ into an element
$\bm{\delta l}$ whose squared norm is
\begin{equation}
\begin{array}{ll}
  \left| \bm{\delta l} \right|^2 &= \bm{\delta l_{0}}^{T}\mathbf{M}^{T}\mathbf{M}\bm{\delta l_{0}}\\ &=\left | \bm{\delta l_{0}} \right |^2 
  \left[e^{2 \lambda t}\cos^2(\psi_{+}-\alpha) + e^{-2 \lambda t}\sin^2(\psi_{+}-\alpha) \right ]
\end{array}
\end{equation}
Noting that the angle $\alpha$ is random, we can show, averaging over $\lambda$, $\psi_{+}$ and $\alpha$, that the total ensemble average length of the contact line is 
(brackets are for an ensemble average):
\begin{equation}\label{E:length_contact}
\begin{array}{ll}
  \langle L \rangle(t)= &\\L_{0} \int_{0}^{\pi} \int_{0}^{\infty} \frac{d\gamma}{\pi} dl P_{\lambda}(t,l) \sqrt{e^{2 l t}\cos^2\gamma +
  e^{-2 l t}\sin^2\gamma} \mbox{,}
\end{array}
\end{equation}
with $L_0$ the initial length of $\mathcal{L}$ and $P_{\lambda}(t,l)$ the time dependent probability density of $\lambda$. Henceforth, the integration bounds over
$l$ and $\gamma$ will be implicit and the same as in ($\ref{E:length_contact}$).  
This expression is valid when the diffusion is not taken into account. 
For large times, when two filaments are brought together at a distance smaller that the diffusive cutoff, they merge under the action 
of diffusion. We think the time span of this regime to be well approximated
by the mix-down time scale $T_{mix} \approx \frac{1}{2 \lambda} \ln \frac{L_e \lambda}{\kappa}$ from the largest scale $L_e$ of the flow to 
the diffusive cutoff (\cite{th97}). If the strain $S$ is an estimation of $\lambda$, which is the case for the flow we study section V and VI 
(figure $\ref{F:STAT}$), and if $\frac{1}{S}$ is an estimate for the integral time scale $T$ of 
the flow, which is also true in our flow, we get that 
\begin{equation}\label{E:mix}
T_{mix} \approx \frac{T}{2} \ln Pe = \frac{T}{2} \ln Re Pr \mbox{,} 
\end{equation}
where $Pe$ is the Peclet number, $Re$ the Reynolds number and $Pr=\frac{Pe}{Re}$ the Prandtl number. 
Our previous work \cite{Ait12} has shown that ($\ref{E:length_contact}$)
is very accurate on time scales of the order of $T_{mix}$ in the flow we describe in section V. 

Assuming the stationarity of the singular vectors, as explained previously, and noting the existence of a local 1-D solution $\chi_t(z)$, as in \cite{bal97}, 
in the direction $z$ normal to the contact line, of the
advection-diffusion equation ($\ref{E:tracer}$) written in the co-moving frame with a contact line element, 
\cite{Ait12} showed that, for an initial gradient profile of tracer $A_0 \delta(z)$, 
with $\delta$ the Dirac delta function, this solution is:
\begin{equation}\label{E:profil}
  \begin{array}{l}
    \chi_t(z)=A_0  \operatorname{Erf} \big( \frac{z G}{2 \sqrt{\kappa}} \big) \\ 
    \mbox { with } G=\sqrt{\frac{e^{2 \lambda t}\cos^2 (\Psi_{+}-\alpha) +e^{-2 \lambda t} \sin^2 (\Psi_{+}-\alpha) }
    {\tau e^{2 \lambda t}\cos^2 (\Psi_{+}-\alpha) +\widetilde{\tau} \sin^2 (\Psi_{+}-\alpha) }}\mbox{.}    
  \end{array}
\end{equation}
The concentration $A_0$ is the initial concentration of reactants $A$ and $B$ in their respective domain. The function $\operatorname{Erf}$ is the gauss error function 
defined on $\mathbb{R}$ as follows: $x \longmapsto \frac{2}{\sqrt{\pi}} \int_0^x e^{-t^2} dt$. The two quantities $\tau$ and $\tilde{\tau}$ are: 
\begin{equation}\label{E:tau} 
    \tau = \frac{\int_0^t e^{2u\lambda(u)}du}{e^{2t\lambda(t)}} \mbox{ and } 
    \widetilde{\tau} = {\int_0^t e^{-2u\lambda(u)}du} \mbox{.} 
\end{equation}
The time $\tau$ has been introduced through the wavenumber growth along Lagrangian trajectories by \cite{Antonsen1996} and was called an equivalent 
time by \cite{Haynes04}. Because the trajectories are chaotic, $\frac{1}{\tau}$ is the stretching rate in the recent past (i.e approximately over the last correlation time).  
\cite{Haynes04} have argued that when the correlation time of the stretching is much smaller than the time scale of the decay of the mean
Lyapunov exponent, $\tau$ becomes independent of $\lambda$ at large times and that its probability distribution converges to a 
time independent form. The time $\widetilde{\tau}$ is also an equivalent time that measures the stretching rate in the early part of the trajectory. As a consequence, we
expect $\tau$ and $\widetilde{\tau}$ to have the same statistics, to be asymptotically equivalent as $t \rightarrow 0$ (typically for times smaller than the
correlation time of the stretching) and to become independent at larger times. 
From $(\ref{E:tau})$, we can see that the gradient along the contact line $\left | \nabla \phi _{\mathcal{L}}\right | \equiv \frac{\partial \chi_t(z)}{\partial z} |_{_{z=0}}$ is:
\begin{equation}\label{E:gradf_general_sol}
  \begin{array}{l}
    \left | \nabla \phi _{\mathcal{L}}\right | = \frac{A_0}{\sqrt{\pi \kappa}} G
  \end{array}
\end{equation}

\section{Probability distribution of the reactant gradients on the contact line}

The distribution of $\left | \nabla \phi _{\mathcal{L}}\right | \frac{\sqrt{\pi \kappa}}{A_0}$ can be inferred form the distribution of $G$  (eq. $\ref{E:gradf_general_sol}$) 
and $\lambda$ and does not depend on $\kappa$. Its probability density function (pdf) along the contact line $P_{G,\mathcal{L}}$ is given by: 
\begin{align}\label{E:pdf_g}
P_{G,\mathcal{L}}(t,g) = \frac{\iint \frac{d\gamma}{\pi} dl P_{G,\lambda}(t,g,l) \sqrt{e^{2 l t}\cos^2\gamma +
e^{-2 l t}\sin^2\gamma }}{ \iint \frac{d\gamma}{\pi} dl P_{\lambda}(t,l) \sqrt{e^{2 l t}\cos^2\gamma +
e^{-2 l t}\sin^2\gamma }}\mbox{,}
\end{align}
where we have introduced the joint pdf $P_{G,\lambda}$ of $G$ and $\lambda$. 
Considering an orbit characterized by $(\lambda,\tau,\widetilde{\tau},\Psi_{+},\alpha)$, expression ($\ref{E:pdf_g}$) can be derived noting that 
$\frac{\sqrt{\pi \kappa}}{A_0}\left | \nabla \phi_{\mathcal L}\right|$ is equal to $G(t,\lambda,\tau,\widetilde{\tau},\Psi_{+},\alpha)$ on a 
fraction of the contact line 
\begin{equation}\label{E:pdf_g_proof}
\frac{\left| \bm{\delta l} \right|/\bm{\left|\delta l_0 \right|}}{\langle L \rangle / L_0}=
\frac{\sqrt{e^{2 \lambda t}\cos^2(\Psi_{+}-\alpha) + e^{-2 \lambda t}\sin^2(\Psi_{+}-\alpha) }}{ \iint \frac{d\gamma}{\pi} dl P_{\lambda}(t,l) \sqrt{e^{2 l t}\cos^2\gamma +
e^{-2 l t}\sin^2\gamma }}\mbox{.}
\end{equation}
For times such that $ t \gg \frac{1}{4 \lambda}$, 
the sine terms in ($\ref{E:pdf_g}$) can be neglected: 
the gradients are equilibrating with the flow and the time
asymptotic form of $G$ is $\frac{1}{\sqrt{\tau}}$ (see ($\ref{E:profil}$)). As a consequence, for 
$ t \gg \frac{1}{4 \lambda} \approx \frac{T}{4}$, $P_{G,\mathcal{L}}(t,g)$ is asymptotically equivalent to $P_{G,\mathcal{L},\infty}$:
\begin{align}\label{E:pdf_g_inf}
P_{G,\mathcal{L}}(t,g) \sim P_{G,\mathcal{L},\infty}(t,g)=\frac{\int dl P_{\frac{1}{\sqrt{\tau}},\lambda}(t,g,l) e^{l t}}
{\int dl P_{\lambda}(t,l) e^{l t}}\mbox{,}
\end{align}
where $P_{\frac{1}{\sqrt{\tau}},\lambda}$ is the joint pdf of $\frac{1}{\sqrt{\tau}}$ and $\lambda$. 
It is worth noting that if $\tau$ and $\lambda$ were independent, which is expected at the very long times, the pdf of $g$ along 
the contact line would be equal to the pdf of $\frac{1}{\sqrt{\tau}}$. This can be seen writing the bivariate density $P_{\frac{1}{\sqrt{\tau}},\lambda}$ as 
the product of its marginal densities in ($\ref{E:pdf_g_inf}$).

\section{Probability distribution of the reactants concentrations}

The reactants are in stoechiometric quantity and their initial concentration in their respective domain is $A_0$. As a consequence, it follows from ($\ref{E:inf_chim}$) that 
the pdfs of $C_A$, $C_B$ and $| \phi |$ are the same. The corresponding random variable will be
noted $\Phi$. Our objective is to understand its dependence with $\kappa$ and 
with time, as well as its shape.  

The profil of the tracer gradient close to the contact line is given by ($\ref{E:profil}$), expression that should be valid as long the contact line is well-defined with a 
curvature larger than $\frac{\sqrt{\kappa}}{g}$, which is expected as long as $t < T_{mix}$. 
We assume that an $\epsilon \ll 1$ can be chosen such that, for all members, the length 
\begin{equation}\label{E:delta_t}
\delta_t=2 \chi_t^{-1} \big(A_0 (1-\epsilon)\big)=\frac{4 \sqrt{\kappa}}{G} \operatorname{Erf}^{-1}(1-\epsilon)
\end{equation}
satisfies $L_a \equiv \sqrt{\mathcal{A}} \gg \delta_t \gg \sqrt{\kappa \tau}$. The range $[-\frac{\delta_t}{2},\frac{\delta_t}{2}]$ is where $|\chi_t|$ takes values smaller than 
$A_0 (1-\epsilon)$. This can be seen noting that $\chi_t$ is a monotonic odd increasing function with  $\chi_t(\pm\frac{\delta_t}{2})=\pm A_0(1-\epsilon)$.
In fact, we have $\frac{L_a}{\sqrt{\kappa \tau}} \ll 1$ because $\frac{L_a}{\sqrt{\frac{\kappa}{S}}} \sim \sqrt{Pe}$ is very large by assumption 
(the Peclet number in the simulations presented in section VI will be of the order of $10^4$ to $10^6$). 
However, $\frac{1}{\tau}$ is in general different form $S$: the former measures a Lagrangian stretching rate in the recent past while the
latter measures an Eulerian stretching rate (for a comparison in the flow considered in section VI, one can refer to figure 1, noting that the strain is the Lyapunov exponent at 
small times). It is possible to choose $\delta_t$ far from both $L_a$ and $\sqrt{\kappa \tau}$ because we are considering a well-defined contact line between areas of 
well mixed reactants. As mentioned earlier, this is expected to be true for $t < T_{mix}$. 

If we consider a profile $ \chi_t $ of tracer around the contact line in the range $[-\frac{\delta_t}{2},\frac{\delta_t}{2}]$, the cumulative 
distribution function (cdf) of $\Phi$, i.e. the probability of having $\Phi$ smaller than a given value $\phi$, is 
\begin{equation}\label{E:cdf_tra_part}
f_{\Phi}(\phi)=\frac{ \chi_t^{-1} (\phi)}{\frac{\delta_t}{2}}=\frac{ \chi_t^{-1} (\phi)}{ \chi_t^{-1} (A_0 (1-\epsilon)}=
\frac{\operatorname{Erf}^{-1}(\frac{\phi}{A_0})}{\operatorname{Erf}^{-1}(1-\epsilon)}
\end{equation}
and stands for values of $\phi$ in the range $[0,A_0 (1-\epsilon)]$.
We multiply by $\frac{\langle\delta_t\rangle  \langle L \rangle }{\mathcal{A}}$ to obtain the cdf $F_{\phi}$ in the whole domain of area $\mathcal{A}$ 
because values in the range  $[0,A_0 (1-\epsilon)]$ are achieved on a sub-domain of area $\langle\delta_t\rangle  \langle L \rangle $:
\begin{equation}\label{E:cdf_tra}
F_{\Phi}(\phi)=\frac{\langle\delta_t\rangle  \langle L \rangle }{\mathcal{A}} \frac{\operatorname{Erf}^{-1}(\frac{\phi}{A_0})}{\operatorname{Erf}^{-1}(1-\epsilon)} \mbox{.}
\end{equation}
Using equation ($\ref{E:delta_t}$) to calculate $\langle\delta_t\rangle$, we have:
\begin{equation}\label{E:cdf_tra_2}
F_{\Phi}(\phi)=\frac{4}{\mathcal{A}}\sqrt{\kappa}  \langle L \rangle  \langle \frac{1}{G}\rangle \operatorname{Erf}^{-1}(\frac{\phi}{A_0}) \mbox{ for } 
\phi \in [0,A_0 (1-\epsilon)] \mbox{.}
\end{equation}
Finally, the probability density is the derivative of the cdf with respect to $\phi$: 
\begin{equation}\label{E:pdf_tra_a}     
       P_{\Phi}(\phi)=\frac{4}{\mathcal{A} A_0} \sqrt{\kappa} \langle L \rangle {\langle \frac{1}{G} \rangle} \operatorname{Erf}^{-1 '} \big(\frac{\phi}{A_0}\big) \mbox{ for } 
       \phi \in [0,A_0 (1-\epsilon)] \mbox{.}  
\end{equation}
A direct consequence of well mixed reactants away from the well-defined contact line is the term $\operatorname{Erf}^{-1 '}$ 
(derivative of the inverse of the Gauss error function). It expresses that the shape of the pdf is an increasing function and can be directly 
related to the profil of the tracer field close to the contact line.
The denisty $P_{\Phi}$ is proportional to $\sqrt{\kappa}{ \langle L \rangle }$ because the area where the field $\phi$ 
takes non-trivial values (i.e significantly different from the initial value $A_0$) is proportional to $\sqrt{\kappa}  \langle L \rangle $: its length is $ \langle L \rangle $ while its width is 
controlled by diffusive processes. Finally the term $\langle \frac{1}{G} \rangle$ depicts the effect of the mean gradient, with a decrease of the gradient along the contact line 
explaining an increase in the probability of small values of $|\phi|$.

\section{Statistics of the Lagrangian stretching properties in a two-dimensional Navier-Stokes flow}

\begin{figure} 
\centering
\includegraphics[scale=0.6]{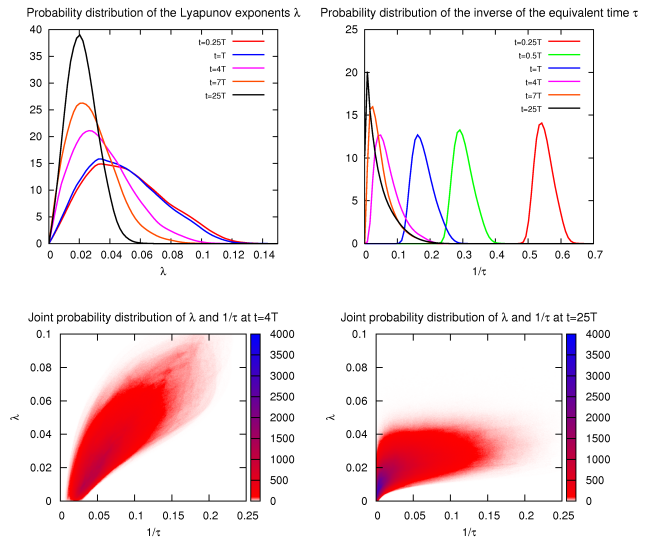}
\caption{\small Probability density of the Lyapunov exponents (top left), of the inverse of the equivalent time $\tau$ defined in ($\ref{E:tau}$) (top right) for 
$0 < t \le 25T$. We note that the density of $\lambda$ at $t=0.25T$ is roughly the density of the strain. 
At the bottom, their joint density at $t=4T$ (left) and $t=25T$ (right).} 
\label{F:STAT}
\end{figure}
The numerical model integrates the following vorticity equation using the pseudo-spectral method:
\begin{equation}\label{E:vorticity}
  \frac{\partial{\omega}}{\partial{t}}+\mathbf{u} \cdot \boldnabla \omega=F-R_0 \omega +  \nu \nabla^2 \omega \mbox{,}
\end{equation}
where $\omega$ is the vorticity, $F$ a forcing term at wavenumber 3, $R_0$ the Rayleigh friction and $\nu$ the viscosity. 
The equation is integrated in a doubly periodic box $[-\pi,\pi] \times [-\pi,\pi]$ on a  $512 \times 512$ grid. The integral time scale
of the flow $T \equiv \sqrt{\frac{2}{\langle \omega^2 \rangle}} \approx \frac{1}{S}$, where the brackets stand for an ensemble average, will be used to normalize the time axis. 
The Reynolds number is of the order of $10^4$.

Trajectories are computed using a fourth order Runge-Kutta scheme with a trilinear interpolation on the velocity field. On each
trajectory, we integrate the resolvent matrix $\mathbf{M}$ such that $\frac{d \mathbf{M}}{dt}= \mathbf{S} \mathbf{M}$ with $\mathbf{S}=\frac{\partial u_j}{\partial x_i}$ the 
velocity gradient tensor along the trajectory and $\mathbf{M}(t=0)$ the identity matrix. The largest eigenvector of the symmetric positive 
matrix $^{t}\mathbf{M}\mathbf{M}$ is $e^{2 \lambda t}$, where $\lambda$ is the Lyapunov exponent on the trajectory at the finite time $t$. 
This method is described in more detail in \cite{Abr02}. We also calculate the equivalent times 
$\tau$ and $\widetilde{\tau}$ 
through a numerical integration of ($\ref{E:tau}$). The trajectories are computed for hundred realizations of the flow, each realization spanning 25 turnover times. 
This gives access to the statistics of $\lambda$, $\tau$, $\widetilde{\tau}$ and $G$ involved in equations ($\ref{E:pdf_g}$) and ($\ref{E:pdf_g_inf}$).

Figure $\ref{F:STAT}$ shows the time evolution of the pdf of $\lambda$ and $\frac{1}{\tau}$. The pdf of $\lambda$ converges to the pdf of the strain $P_s$
as $t \rightarrow 0$ because the strain is the FTLE on each chaotic orbit for an infinitely small time (see ($\ref{E:ftle_def}$)). 
The pdf does not evolve much during the first turnover time, as the correlation time is expected to be of the order of T, or larger. Then, the variance of the 
FTLE decreases while the density shifts toward smaller values. The peak of the density saturates at $\lambda_{max}=0.02$, which we think is a rough estimate
of the infinite time Lyapunov exponent. For a more detailed description of the FTLE, the reader can refer to \cite{ottino02} in chaotic flow, to \cite{Lapeyre02} in 
two-dimensional turbulence and to \cite{Abr02,ngan2,Waugh2008} in geophysical flows. For times smaller than one turnover time, the pdf of $\frac{1}{\tau}$ shifts toward smaller value, 
its shape being only slightly affected. This can be interpreted assuming that $\lambda$ does not evolve much on trajectories and can be estimated by the strain $S$ where the 
trajectory originates. Hence, $\tau \approx 2S/(1-e^{-2 S t})$, which gives $\frac{1}{\tau} \approx \lambda+\frac{1}{t}$ for $t \ll \frac{1}{2S}$. The pdf $P(t,x)$ of 
$\frac{1}{\tau}$ is then approximated by $P_s(x-\frac{1}{t})$ where $P_s$ is the pdf of the strain. In addition, as we expected in section II, we observe that 
the density of $\frac{1}{\tau}$ converges to a time independent form.

Figure $\ref{F:STAT}$ also shows the joint pdf of $\lambda$ and $\frac{1}{\tau}$ at $t=4T$, well within the time range where we think 
($\ref{E:pdf_g}$) should be a satisfying description of the gradient pdf. The joint pdf at $t=25T$ shows this dependence is still important at large times. Previous
studies (e.g \cite{Antonsen1996}) have assumed the independence between $\lambda$ and $\tau$ at long times. This is relevant in simple chaotic flows. However, 
two dimensional Navier-Stokes 
flow exhibit coherent structures (vorticies, filaments of vorticity, etc...) probably making the Lagrangian correlation time dependent on the trajectories. 
In particular, very long correlation times could be associated with trajectories trapped in vorticies, where the stretching rate is particularly weak. This could explain 
the strong dependence between large values of $\tau$ and small values of $\lambda$. 

\section{Numerical results}

\subsection{Gradients along the contact line}

\begin{figure*} 
\centering
\includegraphics[scale=0.7]{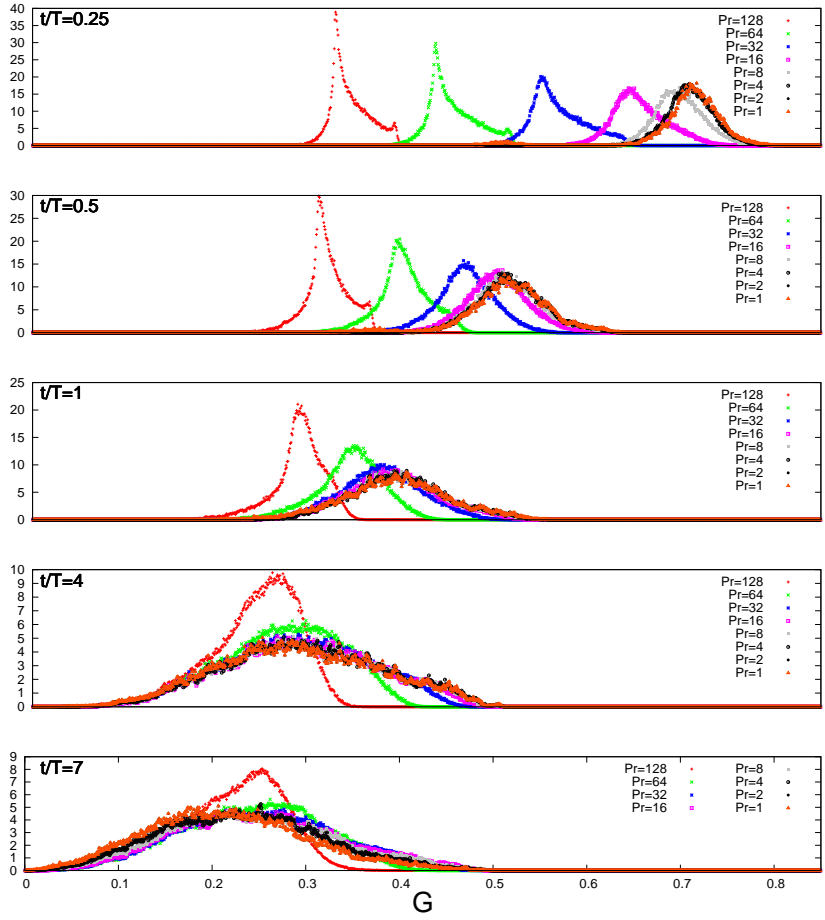}
\caption{\small Probability density  $P_{G,1 \le Pr \le 128}$ of $G_{e,Pr}=\frac{\sqrt{\pi \kappa}}{A_0} |\nabla \phi_{\mathcal{L}}|$ where $|\nabla \phi_{\mathcal{L}}|$ is
the modulus of the gradient of $\phi$ along the line $\mathcal{L}=\{\mathbf{x}\lvert\phi(\mathbf{x})=0\}$. These pdf are obtained from an ensemble of 34 direct numerical 
simulations and plotted here for $t=\frac{1}{4}T,\frac{1}{2}T,T,4T \mbox{ and } 7T$} 
\label{F:GRAD_CONTACT_PR_TIME}
\end{figure*}

The numerical simulations are performed for eight different Prandtl numbers 
$Pr\equiv \frac{\kappa}{\nu}=2^i \mbox{ for } 0 \leq i \leq 7$. For each one we run an ensemble of 34 simulations 
integrating equations ($\ref{E:tracer}$) and ($\ref{E:vorticity}$) in the periodic box. Each
member is defined by its initial condition on the flow, taken as the vorticity field every turnover time of a long time simulation of the 
statistically stationary flow solution of ($\ref{E:vorticity}$). For each member, we use the following initial condition on the tracer:
\begin{equation} \label{E:init-trac}
    \phi(x,y,t=0)=2 A_0 (H(x)-\frac{1}{2}) \mbox{  }\mbox{ for }(x,y)\in[-\pi,\pi]^2 \mbox{,}   
\end{equation}
where $H$ is the Heaviside step function. In other words, in one half of the box, $C_A=A_0$ and $C_B=0$, and in the other half 
$C_A=0$, and $C_B=A_0$. $A$ and $B$ are separated by initially infinite gradients. 
This is not exactly true in the numerical integrations because of the finite resolution of the
model and has to be kept in mind for an accurate interpretation of the numerical results.

For each member we determine the coordinates of the contact line 
$\mathcal{L}$ with a time increment $\frac{T}{4}$ using the library DISLIN (\cite{dislin}). We calculate the modulus of the gradient of $\phi$ at each of these 
coordinates using bilinear interpolation. We then multiply it by $\frac{\sqrt{\pi \kappa}}{A_0}$ in order to obtain a 
physical quantity that scales like $G$ in equation ($\ref{E:profil}$) and which we name $G_{e,Pr}$.
A weighted histogram of $G_{e,Pr}$ is calculated every time increment $\frac{T}{4}$ using as weight the length enclosed by three consecutive points of  
$\mathcal{L}$ centered in the point where $G_{e,Pr}$ is estimated. This is necessary since the points are not equidistant. The probability 
densities obtained after normalization of these histograms are noted $P_{G,Pr}$.

On figure $\ref{F:GRAD_CONTACT_PR_TIME}$ we plot  $P_{G,Pr}$ for different Prandtl numbers and different times. We observe:

\begin{itemize}
       \item When time becomes shorter, as observed from $t=4T$ to $t=0.25T$, the independence of $G_{e,Pr}$ to $Pr$, as predicted in sections III and IV, is not verified.
	     As a matter of fact, the gradient cannot be considered as infinite at the initial time because of the finite grid size of the numerical model. 
	     This effect can be quantified: we can solve the advection diffusion equation
	     in a Lagrangian co-moving frame with a contact line element for the tracer profile around this line with the initial condition on the gradients 
	     $\frac{\partial \chi_t}{\partial z}|_{_{t=0}}
	     =\frac{A_0}{2 \delta_0 \sqrt{\pi}} e^{-\frac{x^2}{4 \delta_0^2}}$ ($\delta_0$ is a length corresponding to a grid point). We get:
	      \begin{equation}\label{E:gradf_general_sol_b}
		\begin{array}{l}
		  \left | \nabla \phi_{\mathcal{L}}\right| = \frac{A_0}{\sqrt{\pi \kappa}} G_{\kappa}  \\
		  \mbox { with } G_{\kappa}= \sqrt{\frac{e^{2 \lambda t}\cos^2 (\Psi_{+}-\alpha +e^{-2 \lambda t} \sin^2 (\Psi_{+}-\alpha) }
		  {\frac{\delta_0^2}{\kappa}+[\tau e^{2 \lambda t}\cos^2 (\Psi_{+}-\alpha) +\widetilde{\tau} \sin^2 (\Psi_{+}-\alpha )]}}\mbox{,}
		\end{array}
	      \end{equation}
	    where $G_{\kappa}$ now depends on diffusion through the term $\frac{\delta_0^2}{\kappa}$ (this expression has to be compared to $G$ in ($\ref{E:profil}$)). 
	    The initial gradient cannot be assumed infinite when the diffusive 
	    cutoff $\sqrt{\kappa \tau}$ is of the order of the grid size $\delta_0$. The time scale $T_{Pr}$ 
	    where this effect is important can be evaluated by comparing the two terms $\frac{\delta_0^2}{\kappa}$ and 
	    $[\tau e^{2 \lambda t}\cos^2 (\Psi_{+}-\alpha) +\widetilde{\tau} \sin^2 (\Psi_{+}-\alpha )]$ in the denominator of $G_{\kappa}^2$. 
	    Approximating $\lambda$ with $S$, we find $T_{Pr} \approx \frac{1}{2S}\ln(1+\frac{2\delta_0^2}{\frac{\nu}{S}}Pr)$, which gives respectively
	    $T_{Pr}/T=0.12;0.25;0.45;0.75;1.2;1.6$ for $Pr=4;8;16;32;64;128$. This calculation is consistent with the numerical results presented on figure
	    $\ref{F:GRAD_CONTACT_PR_TIME}$, the pdf $P_{G,Pr}$ being independent of $Pr$ for $t \gtrsim T_{Pr}$

      \item At $t=4T$, the densities $P_{G,Pr=128}$ and $P_{G,Pr=64}$ are different from $P_{G,{Pr \le 32}}$. Because of the finite resolution of the model, gradients at the high
	    end of these distributions cannot be resolved. This explains the strong asymmetry of these densities. This effect is also observed at other times: in particular, 
	    it explains the kick at the very high end of the densities $G_{Pr \ge 32}$ at $t=0.25T$. 

      \item At $t=7T$, the densities $P_{G,Pr}$ are not anymore independent of $Pr$ because we are at $t\gtrsim  T_{mix}(\kappa)$ for all the Prandtl numbers
\end{itemize}
 
\begin{figure*} 
\centering
\includegraphics[scale=0.7]{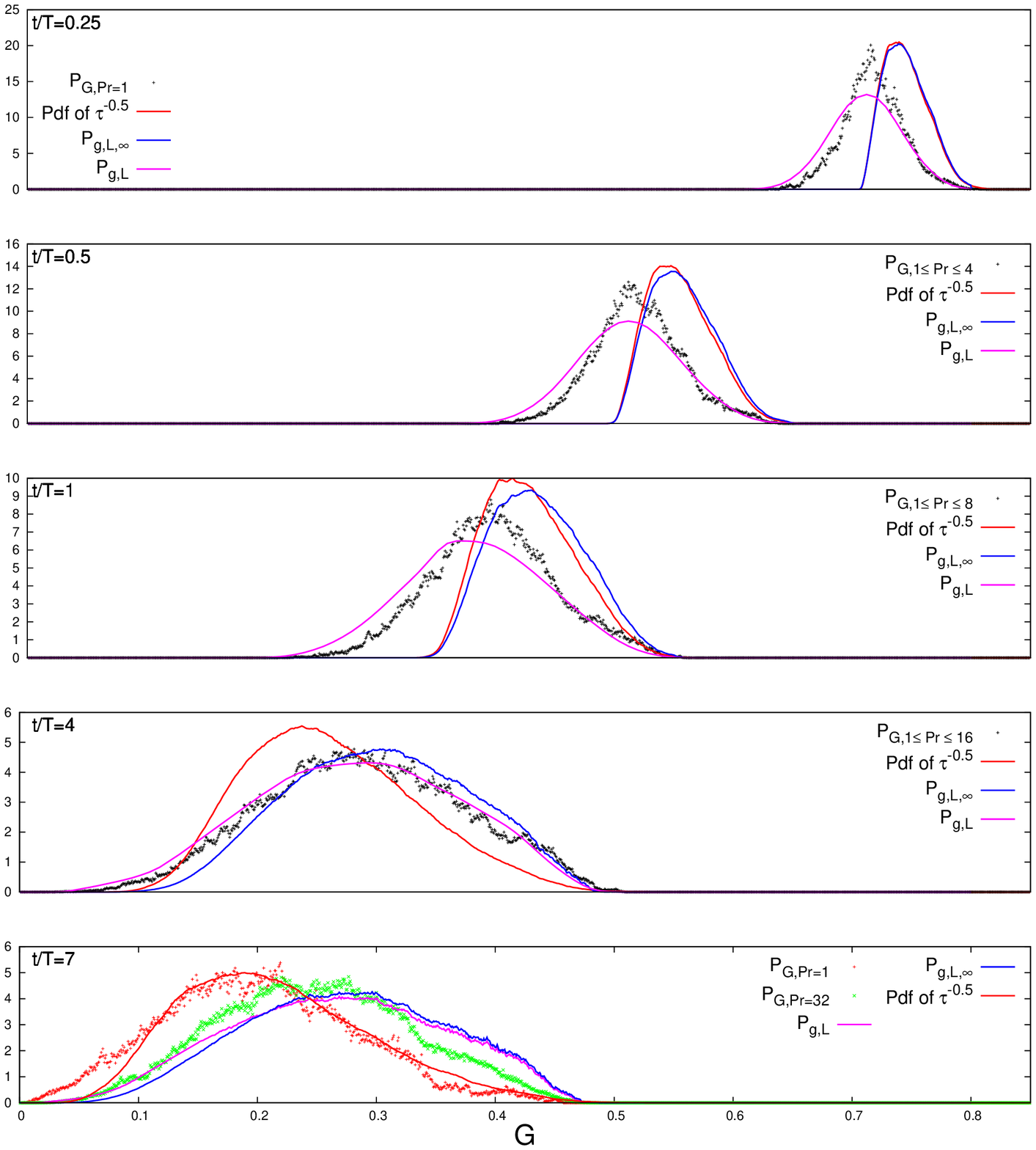}
\caption{\small Comparison between $P_{G,Pr}$ obtained from the direct numerical simulations and the theoretical predictions $P_{G,\mathcal{L}}$ (eq. $\ref{E:pdf_g}$), 
$P_{G,\mathcal{L},\infty}$ (eq. $\ref{E:pdf_g_inf}$) and the pdf of $\frac{1}{\sqrt{\tau}}$ ($\tau$ is defined in ($\ref{E:tau}$)), obtained from the calculation of
the Lagrangian stretching properties of the flow calculated with the trajectories. We have only plotted the curves $P_{G,Pr}$ corresponding to direct numerical simulations 
consistent with the infinite initial gradient hypothesis.} 
\label{F:GRAD_CONTACT_COMPAR_TH}
\end{figure*}

In order to remain consistent with the infinite initial gradient hypothesis, we will only consider small enough Prandtl numbers.
We compare on figure $\ref{F:GRAD_CONTACT_COMPAR_TH}$ the numerical results to the theoretical predictions $P_{G,\mathcal{L}}$ 
($\ref{E:pdf_g}$), $P_{G,\mathcal{L},\infty}$ ($\ref{E:pdf_g_inf}$), and to the pdf of $\frac{1}{\sqrt{\tau}}$.
\begin{itemize}
  \item For $t \lesssim T$, only $P_{G,\mathcal{L}}$ shows some 
	success in predicting $P_{G,Pr}$. This is consistent with ($\ref{E:pdf_g}$) which states that the effect of 
	$\gamma$ cannot be neglected at times smaller than $T$. 

  \item For $t=4T$,  $P_{G,\mathcal{L}}$ and $P_{G,\mathcal{L},\infty}$ are much more similar because the contact line elements have equilibrated with the flow (their orientation
	does not depend anymore on their initial orientation $\alpha$). 
	The agreement between $P_{G,\mathcal{L}}$ and  $P_{G,Pr}$ is very good. 

  \item For $t=7T$, $P_{G,\mathcal{L}}$ and $P_{G,\mathcal{L},\infty}$ are even closer. $P_{G,\mathcal{L},\infty}$ fails to predict $P_{G,Pr=1}$ 
	but performs reasonably for $P_{G,Pr=32}$. Actually, an estimate of $T_{mix}$ from ($\ref{E:mix}$) gives $4.5T$ for $Pr=1$ and $6.5T$ for $Pr=32$, which is 
	consistent with the discrepancy between $P_{G,Pr=1}$ and $P_{G,Pr=32}$ at $t=7T$.
\end{itemize}

It is worth observing that the pdf of $\frac{1}{\sqrt{\tau}}$ fails to predict $P_{G,Pr}$ because of the dependence of $\lambda$ with $\tau$, 
which is significant even at very long times (section V)\footnote[1]{The similarity between 
the pdf of $\frac{1}{\sqrt{\tau}}$ and $P_{G,Pr=1}(t=7T)$ (figure $\ref{F:GRAD_CONTACT_COMPAR_TH}$) is not explained by our theory, as showed by the significant 
difference between the pdf of $\frac{1}{\sqrt{\tau}}$ and $P_{G,\mathcal{L}}$.}. We observe that the time scale for the much simpler $P_{G,\mathcal{L},\infty}$ to become a good 
prediction for the gradient, i.e the timescale for $P_{G,\mathcal{L}}$ to converge to $P_{G,\mathcal{L},\infty}$, seems to be of the order of $T_{mix}$. This non-trivial behavior 
may be determined by the dependence of $\tau$ with $\widetilde{\tau}$ ($\ref{E:tau}$). If we had $\tau \sim \widetilde{\tau}$, which is expected as $t \rightarrow 0$, 
this convergence would have been of the order of $\frac{1}{2 \lambda} \approx \frac{1}{2 S} \approx \frac{T}{2}$ (equations ($\ref{E:gradf_general_sol}$) and ($\ref{E:pdf_g}$)). 
It is one order of magnitude longer. The reason could  lie in the fact that $\widetilde{\tau}$ and  $\tau$ quickly become independent.

Equation ($\ref{E:pdf_g}$), where the joint pdf of $\lambda$ and $G$ is replaced by the joint pdf of $\lambda$ and $G_{\kappa}$, is a fair approximation to 
$P_{G,Pr \le 32}$ at any time (not shown). This extends our results to finite initial gradients, but concentrated on length scales not large compared to the
diffusive cutoff of the flow, such that a Lagrangian straining theory approach remains possible.

\subsection{Reactants' fields}

\begin{figure} 
\centering
\includegraphics[scale=0.65]{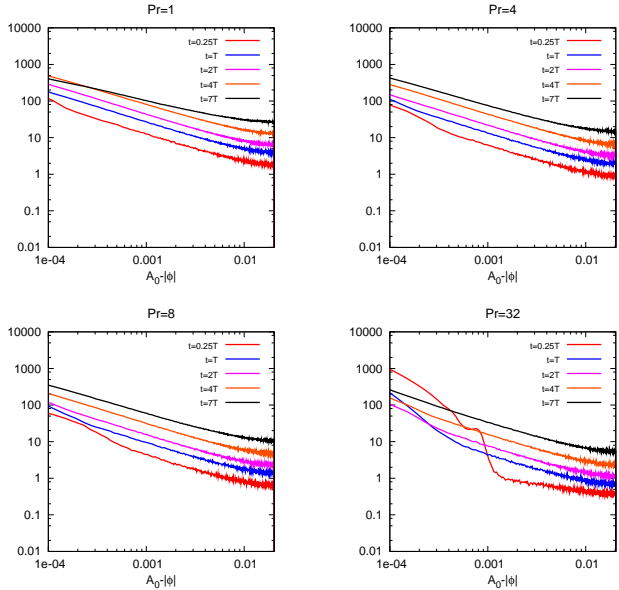}
\caption{\small Numerically determined pdf $Q_{Pr}$ of $\widetilde{\phi} \equiv A_0-|\phi|$ at different times $t=\frac{1}{4}T,T,2T,4T \mbox{ and } 7T$ 
and for Prandtl numbers $Pr=1,4,8\mbox{ and }32$. Log-log scale.} 
\label{F:PDF}
\end{figure}

We calculate the pdf $Q_{Pr}$ of $\widetilde{\phi} \equiv A_0-|\phi|$ using our 34 simulations ensemble for the entire range of Prandtl numbers.
On figure $\ref{F:PDF}$, we show $Q_{Pr}$ for $Pr=1,4,8 \mbox{ and } 32$ and for $0.25T \le t \le 7T$. Our objective is to 
compare the theoretical prediction ($\ref{E:pdf_tra_a}$) to these numerical results. 

\begin{figure} 
\centering
\includegraphics[scale=0.65]{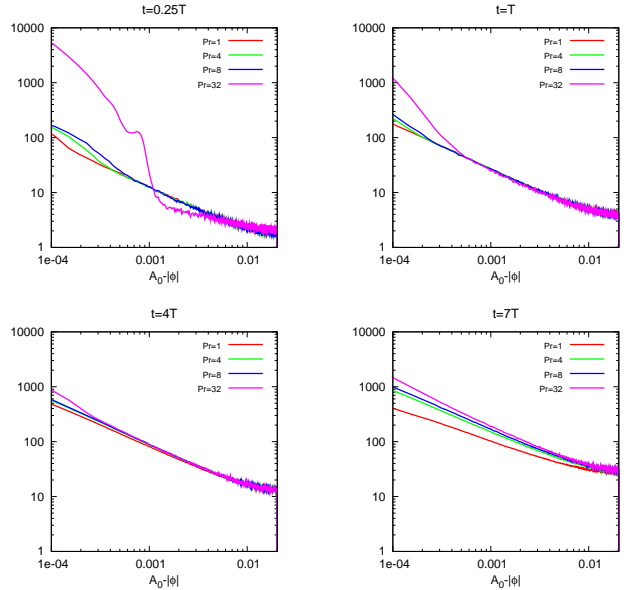}
\caption{\small Numerically determined pdf $Q_{Pr}$ of $
\widetilde{\phi} \equiv A_0-|\phi|$ multiplied by $\sqrt{Pr}$ for $Pr=1,4,8\mbox{ and }32$ 
and $t=\frac{1}{4}T,T,t=4T \mbox{ and } 7T$. Log-log scale.} 
\label{F:PDF_DIFF}
\end{figure}

Figure $\ref{F:PDF_DIFF}$ shows $\sqrt{Pr} Q_{Pr}$ for a wide range of Prandtl numbers over the first few turnover times. The dependence in $\sqrt{\kappa}$
predicted in ($\ref{E:pdf_tra_a}$) is well achieved, except at small times ($t=0.25T$) especially for small diffusion ($Pr=32$), i.e. when the infinite gradient
assumption is again violated in the numerical simulations. It also fails at $t=7T$ because $T > T_{mix}$. 

\begin{figure} 
\centering
\includegraphics[scale=1.3]{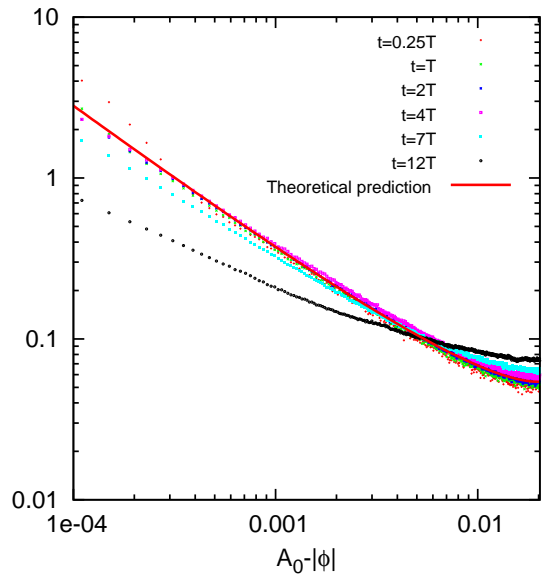}
\caption{\small Time evolution of $Q_{Pr=8} \sqrt{Pr} \frac{\langle 1/G \rangle}{\langle L \rangle}$. $ \langle L \rangle$ and $G$ are defined respectively in equations 
($\ref{E:length_contact}$) and ($\ref{E:profil}$) and are calculated from trajectories, as described in section V. 
The red curve (theoretical prediction) corresponds $\frac{4\sqrt{\nu}}{\mathcal{A} A_0} \operatorname{Erf}^{-1 '} \big(\frac{A_0-\tilde{\phi}}{A_0}\big)$, 
where $\operatorname{Erf}$ is the Gauss error function. Log-log scale.} 
\label{F:PDF_TEMPS}
\end{figure}

Next, we compare the time dependence predicted in ($\ref{E:pdf_tra_a}$) to the numerical results. 
For $Pr=8$, we show the product of $Q_{Pr}$ with $\sqrt{Pr} \frac{\langle 1/G \rangle}{\langle L \rangle}$ on figure $\ref{F:PDF_TEMPS}$. 
The contact line length $\langle L \rangle$ is calculated form $P_\lambda$ using ($\ref{E:length_contact}$), and $\langle 1/G \rangle$ from the 
integral $\int_0^{\infty} \frac{1}{g} P_{G,\mathcal{L}}(t,g) dg$ where $P_{G,\mathcal{L}}$ is defined in ($\ref{E:pdf_g}$). The densities $P_\lambda$ and $P_{G,\mathcal{L}}$ are 
determined as explained in section V from the computation of Lagrangian trajectories. 
The curves converges together, except at very small times
for values of $|\phi|$ close to $A_0$ ($\widetilde{\phi}$ close to 0). As expected, ($\ref{E:pdf_tra_a}$) does not work either for $t> T_{mix}$, as shown by the curves
$t=7T$ and $t=12T$.

The shape of the reactants' pdf shown as $Q_{Pr=8} \sqrt{Pr} \frac{\langle 1/G \rangle}{\langle L \rangle}$ on figure $\ref{F:PDF_TEMPS}$ is very well reproduced by 
$\frac{4\nu}{\mathcal{A} A_0} \operatorname{Erf}^{-1 '} \big(\frac{A_0-\tilde{\phi}}{A_0}\big)$, which confirms the theoretical prediction ($\ref{E:pdf_tra_a}$).

\section{Conclusion and discussion}

We have considered the early regime of an instantaneous chemical reaction in a two-dimensional Navier-Stokes flow, for segregated reactants 
initially separated by infinite gradients. The time scales considered here are shorter than the mix-down time scale from the integral length scale to the diffusive cutoff 
($T_{mix}$ given in ($\ref{E:mix}$)). Assuming that the singular vector associated with a FTLE does not depend on time and is equal to the forward Lyapunov vector,
we have adopted a Lagrangian straining theory approach and showed numerically its success in predicting (a) the statistics of
the diffusive flux of fast reacting chemicals along their interface ($\ref{E:pdf_g}$), and thus the chemical speed, (b) the probability distribution of the
reactants ($\ref{E:pdf_tra_a}$). We have put the emphasis on the effect of the
reactants’ diffusion, showing (a) that the distribution of
the gradients along the contact line, rescaled by
$\kappa^{\frac{1}{2}}$,
does not depend on $\kappa$ (b) that the distribution of the reactant concentration, when it is not too close to $A_0$, is proportional to $\kappa^{\frac{1}{2}}$.

At the very early stage, for about one turnover time, predicting these statistics requires
to know the joint statistics of $G(\lambda, \tau, \widetilde{\tau}, \gamma \equiv \Psi_{+}-\alpha)$ (equation $\ref{E:profil}$) and $\lambda$, 
which is given by the joint statistics of $\lambda$, $\tau$ and $\widetilde{\tau}$ ($\gamma$ is a random angle because $\alpha$ is a random angle, and is thus independent of the 
other variables). At moderate time scales 
(a few turnover times), the knowledge of the joint pdf of $\lambda$ and $\tau$ is sufficient. More work is needed to understand these distributions, i.e the marginal 
distributions of $\lambda$, $\tau$ and $\widetilde{\tau}$ and how these three variables depend on each other. 

Previous studies (\cite{Waugh2006,Waugh2008}) investigating the two-dimensional 
mixing in the upper layer of the ocean suggest that it 
might be possible to recover the distribution of $\lambda$ from the distribution of the strain (which is $\lambda$ as $t \rightarrow 0$), 
a more accessible Eulerian quantity, and from the time 
evolution of the mean Lyapunov exponent $\langle \lambda \rangle$. Specifically, they observed that the distribution of
$\frac{\lambda}{\langle \lambda \rangle}$ does not change significantly in time over several turnover times
and is given by a Weibull distribution. We have observed the same behavior in our flow (not shown), with the distribution of the strain being very well approximated by a Weibull 
distribution of shape parameter $1.8$\footnote[1]{This is very close to a Rayleigh distribution (Weibull distribution of shape parameter $2$), which is the distribution 
of the norm of a vector whose components are independent from each other and follow Gaussian statistics}. The time evolution of $\langle \lambda \rangle$ in a chaotic ergodic 
flow was theoretically predicted by \cite{Tang1996}, a prediction which was shown to perform well in the mixing layer of the ocean (\cite{Abr02}). 
Nevertheless, the mechanisms in two-dimensional Navier-Stokes flows (or in similar geophysical flows) behind the evolution of the finite time Lyapunov exponent distribution
toward smaller values (figure $\ref{F:STAT}$) were not addressed in the literature to our knowledge. Although this aspect deserves more investigation, we can speculate that
parcels tend to stay longer in area of low strain than in areas of high strain. For example, a
parcel can be captured in a vortex for a long time but travels very quickly in areas of high
strain. Hence we would have two explanations for this shift: (a) the fact that regions with
low strain (e.g vorticies) have smaller velocities than regions of high strain; (b) the existence
of barriers of transport at the edge of the vorticies that act like parcel traps.
 
We have shown that the distribution of $\tau$ can be obtained from the strain distribution
below one turnover time. Its time evolution at longer time scale has to be studied in more detail, especially
its asymptotic form. The dependence between $\tau$ and $\lambda$ is a complex issue which needs to be addressed. We expect it to be strongly 
dependent on the nature of the flow, especially on its Lagrangian correlation time. In our two-dimensional Navier-Stokes flow, areas of low stretching associated with long 
correlation times (e.g. vorticies) could explain the strong dependence between $\lambda$ and $\frac{1}{\tau}$ where they are both small compared to their ensemble mean (figure 
$\ref{F:STAT}$).

Some studies (e.g. \cite{Fereday04,Haynes05}) have shown, both numerically and theoretically, the relevance of
Lagrangian stretching theories when the length scale of the tracer are much smaller than the scale of the flow, like in
the present study. However, this was done for simple prescribed smooth chaotic flows and in
the long time decay. This theory was recently applied by \cite{tsang09} to infinite chemistry in the long
time decay. Here we apply it for the initial regime. A comprehensive picture would be given
by studying intermediate time scales, which will be the subject of a future paper.

\bibliography{lit_sav.bib}

\providecommand{\noopsort}[1]{}\providecommand{\singleletter}[1]{#1}%
\begin{thebibliography}{21}%
\makeatletter
\providecommand \@ifxundefined [1]{%
 \@ifx{#1\undefined}
}%
\providecommand \@ifnum [1]{%
 \ifnum #1\expandafter \@firstoftwo
 \else \expandafter \@secondoftwo
 \fi
}%
\providecommand \@ifx [1]{%
 \ifx #1\expandafter \@firstoftwo
 \else \expandafter \@secondoftwo
 \fi
}%
\providecommand \natexlab [1]{#1}%
\providecommand \enquote  [1]{``#1''}%
\providecommand \bibnamefont  [1]{#1}%
\providecommand \bibfnamefont [1]{#1}%
\providecommand \citenamefont [1]{#1}%
\providecommand \href@noop [0]{\@secondoftwo}%
\providecommand \href [0]{\begingroup \@sanitize@url \@href}%
\providecommand \@href[1]{\@@startlink{#1}\@@href}%
\providecommand \@@href[1]{\endgroup#1\@@endlink}%
\providecommand \@sanitize@url [0]{\catcode `\\12\catcode `\$12\catcode
  `\&12\catcode `\#12\catcode `\^12\catcode `\_12\catcode `\%12\relax}%
\providecommand \@@startlink[1]{}%
\providecommand \@@endlink[0]{}%
\providecommand \url  [0]{\begingroup\@sanitize@url \@url }%
\providecommand \@url [1]{\endgroup\@href {#1}{\urlprefix }}%
\providecommand \urlprefix  [0]{URL }%
\providecommand \Eprint [0]{\href }%
\providecommand \doibase [0]{http://dx.doi.org/}%
\providecommand \selectlanguage [0]{\@gobble}%
\providecommand \bibinfo  [0]{\@secondoftwo}%
\providecommand \bibfield  [0]{\@secondoftwo}%
\providecommand \translation [1]{[#1]}%
\providecommand \BibitemOpen [0]{}%
\providecommand \bibitemStop [0]{}%
\providecommand \bibitemNoStop [0]{.\EOS\space}%
\providecommand \EOS [0]{\spacefactor3000\relax}%
\providecommand \BibitemShut  [1]{\csname bibitem#1\endcsname}%
\let\auto@bib@innerbib\@empty
\bibitem [{\citenamefont {Tan}\ \emph {et~al.}(1998)\citenamefont {Tan},
  \citenamefont {Haynes}, \citenamefont {MacKenzie},\ and\ \citenamefont
  {Pyle}}]{Tan1998}%
  \BibitemOpen
  \bibfield  {author} {\bibinfo {author} {\bibfnamefont {D.}~\bibnamefont
  {Tan}}, \bibinfo {author} {\bibfnamefont {P.}~\bibnamefont {Haynes}},
  \bibinfo {author} {\bibfnamefont {A.}~\bibnamefont {MacKenzie}}, \ and\
  \bibinfo {author} {\bibfnamefont {J.}~\bibnamefont {Pyle}},\ }\href@noop {}
  {\bibfield  {journal} {\bibinfo  {journal} {Journal of Geophysical
  Research-Atmospheres}\ }\textbf {\bibinfo {volume} {103}},\ \bibinfo {pages}
  {1585} (\bibinfo {year} {1998})}\BibitemShut {NoStop}%
\bibitem [{\citenamefont {Edouard}\ \emph {et~al.}(1996)\citenamefont
  {Edouard}, \citenamefont {Legras}, \citenamefont {Lefevre},\ and\
  \citenamefont {Eymard}}]{Edouard1996}%
  \BibitemOpen
  \bibfield  {author} {\bibinfo {author} {\bibfnamefont {S.}~\bibnamefont
  {Edouard}}, \bibinfo {author} {\bibfnamefont {B.}~\bibnamefont {Legras}},
  \bibinfo {author} {\bibfnamefont {F.}~\bibnamefont {Lefevre}}, \ and\
  \bibinfo {author} {\bibfnamefont {R.}~\bibnamefont {Eymard}},\ }\href@noop {}
  {\bibfield  {journal} {\bibinfo  {journal} {Nature}\ }\textbf {\bibinfo
  {volume} {384}},\ \bibinfo {pages} {444} (\bibinfo {year}
  {1996})}\BibitemShut {NoStop}%
\bibitem [{\citenamefont {Wonhas}\ and\ \citenamefont
  {Vassilicos}(2002)}]{Wonhas02}%
  \BibitemOpen
  \bibfield  {author} {\bibinfo {author} {\bibfnamefont {A.}~\bibnamefont
  {Wonhas}}\ and\ \bibinfo {author} {\bibfnamefont {J.}~\bibnamefont
  {Vassilicos}},\ }\href@noop {} {\bibfield  {journal} {\bibinfo  {journal}
  {Physical Review E}\ }\textbf {\bibinfo {volume} {65}},\ \bibinfo {pages}
  {051111} (\bibinfo {year} {2002})}\BibitemShut {NoStop}%
\bibitem [{\citenamefont {Ait-Chaalal}\ \emph {et~al.}(view)\citenamefont
  {Ait-Chaalal}, \citenamefont {Bourqui},\ and\ \citenamefont
  {Bartello}}]{Ait12}%
  \BibitemOpen
  \bibfield  {author} {\bibinfo {author} {\bibfnamefont {F.}~\bibnamefont
  {Ait-Chaalal}}, \bibinfo {author} {\bibfnamefont {M.~S.}\ \bibnamefont
  {Bourqui}}, \ and\ \bibinfo {author} {\bibfnamefont {P.}~\bibnamefont
  {Bartello}},\ }\href@noop {} {\bibfield  {journal} {\bibinfo  {journal}
  {Physical Review E}\ } (\bibinfo {year} {under review})}\BibitemShut
  {NoStop}%
\bibitem [{\citenamefont {Oseledec}(1968)}]{Os68}%
  \BibitemOpen
  \bibfield  {author} {\bibinfo {author} {\bibfnamefont {V.~I.}\ \bibnamefont
  {Oseledec}},\ }\href@noop {} {\bibfield  {journal} {\bibinfo  {journal}
  {Trans. Moscow Math Soc.}\ }\textbf {\bibinfo {volume} {19}},\ \bibinfo
  {pages} {197} (\bibinfo {year} {1968})}\BibitemShut {NoStop}%
\bibitem [{\citenamefont {Tang}\ and\ \citenamefont {Boozer}(1996)}]{Tang1996}%
  \BibitemOpen
  \bibfield  {author} {\bibinfo {author} {\bibfnamefont {X.}~\bibnamefont
  {Tang}}\ and\ \bibinfo {author} {\bibfnamefont {A.}~\bibnamefont {Boozer}},\
  }\href@noop {} {\bibfield  {journal} {\bibinfo  {journal} {Physica D}\
  }\textbf {\bibinfo {volume} {95}},\ \bibinfo {pages} {283} (\bibinfo {year}
  {1996})}\BibitemShut {NoStop}%
\bibitem [{\citenamefont {Goldhirsch}\ \emph {et~al.}(1987)\citenamefont
  {Goldhirsch}, \citenamefont {Sulem},\ and\ \citenamefont {Orsazg}}]{Gol87}%
  \BibitemOpen
  \bibfield  {author} {\bibinfo {author} {\bibfnamefont {I.}~\bibnamefont
  {Goldhirsch}}, \bibinfo {author} {\bibfnamefont {P.}~\bibnamefont {Sulem}}, \
  and\ \bibinfo {author} {\bibfnamefont {S.}~\bibnamefont {Orsazg}},\
  }\href@noop {} {\bibfield  {journal} {\bibinfo  {journal} {Physica D}\
  }\textbf {\bibinfo {volume} {27}},\ \bibinfo {pages} {311} (\bibinfo {year}
  {1987})}\BibitemShut {NoStop}%
\bibitem [{\citenamefont {Lapeyre}(2002)}]{Lapeyre02}%
  \BibitemOpen
  \bibfield  {author} {\bibinfo {author} {\bibfnamefont {G.}~\bibnamefont
  {Lapeyre}},\ }\href@noop {} {\bibfield  {journal} {\bibinfo  {journal}
  {Chaos}\ }\textbf {\bibinfo {volume} {12}},\ \bibinfo {pages} {688} (\bibinfo
  {year} {2002})}\BibitemShut {NoStop}%
\bibitem [{\citenamefont {Thuburn}\ and\ \citenamefont {Tan}(1997)}]{th97}%
  \BibitemOpen
  \bibfield  {author} {\bibinfo {author} {\bibfnamefont {J.}~\bibnamefont
  {Thuburn}}\ and\ \bibinfo {author} {\bibfnamefont {D.}~\bibnamefont {Tan}},\
  }\href@noop {} {\bibfield  {journal} {\bibinfo  {journal} {Journal of
  Geophysical Research-Atmospheres}\ }\textbf {\bibinfo {volume} {102}},\
  \bibinfo {pages} {13037} (\bibinfo {year} {1997})}\BibitemShut {NoStop}%
\bibitem [{\citenamefont {Balluch}\ and\ \citenamefont {Haynes}(1997)}]{bal97}%
  \BibitemOpen
  \bibfield  {author} {\bibinfo {author} {\bibfnamefont {M.}~\bibnamefont
  {Balluch}}\ and\ \bibinfo {author} {\bibfnamefont {P.}~\bibnamefont
  {Haynes}},\ }\href@noop {} {\bibfield  {journal} {\bibinfo  {journal}
  {Journal of Geophysical Research-Atmospheres}\ }\textbf {\bibinfo {volume}
  {102}},\ \bibinfo {pages} {23487} (\bibinfo {year} {1997})}\BibitemShut
  {NoStop}%
\bibitem [{\citenamefont {Antonsen}\ \emph {et~al.}(1996)\citenamefont
  {Antonsen}, \citenamefont {Fan}, \citenamefont {Ott},\ and\ \citenamefont
  {Garcia~Lopez}}]{Antonsen1996}%
  \BibitemOpen
  \bibfield  {author} {\bibinfo {author} {\bibfnamefont {T.~M.}\ \bibnamefont
  {Antonsen}}, \bibinfo {author} {\bibfnamefont {Z.~C.}\ \bibnamefont {Fan}},
  \bibinfo {author} {\bibfnamefont {E.}~\bibnamefont {Ott}}, \ and\ \bibinfo
  {author} {\bibfnamefont {E.}~\bibnamefont {Garcia~Lopez}},\ }\href@noop {}
  {\bibfield  {journal} {\bibinfo  {journal} {Physics of Fluids}\ }\textbf
  {\bibinfo {volume} {8}},\ \bibinfo {pages} {3094} (\bibinfo {year}
  {1996})}\BibitemShut {NoStop}%
\bibitem [{\citenamefont {Haynes}\ and\ \citenamefont
  {Vanneste}(2004)}]{Haynes04}%
  \BibitemOpen
  \bibfield  {author} {\bibinfo {author} {\bibfnamefont {P.}~\bibnamefont
  {Haynes}}\ and\ \bibinfo {author} {\bibfnamefont {J.}~\bibnamefont
  {Vanneste}},\ }\href@noop {} {\bibfield  {journal} {\bibinfo  {journal}
  {Journal of the Atmospheric Sciences}\ }\textbf {\bibinfo {volume} {61}},\
  \bibinfo {pages} {161} (\bibinfo {year} {2004})}\BibitemShut {NoStop}%
\bibitem [{\citenamefont {Abraham}\ and\ \citenamefont {Bowen}(2002)}]{Abr02}%
  \BibitemOpen
  \bibfield  {author} {\bibinfo {author} {\bibfnamefont {E.~R.}\ \bibnamefont
  {Abraham}}\ and\ \bibinfo {author} {\bibfnamefont {M.~M.}\ \bibnamefont
  {Bowen}},\ }\href@noop {} {\bibfield  {journal} {\bibinfo  {journal} {Chaos}\
  }\textbf {\bibinfo {volume} {12}},\ \bibinfo {pages} {373} (\bibinfo {year}
  {2002})}\BibitemShut {NoStop}%
\bibitem [{\citenamefont {Ott}(2002)}]{ottino02}%
  \BibitemOpen
  \bibfield  {author} {\bibinfo {author} {\bibfnamefont {R.}~\bibnamefont
  {Ott}},\ }\href@noop {} {\emph {\bibinfo {title} {Chaos in Dynamical
  Systems}}}\ (\bibinfo  {publisher} {Cambridge, England},\ \bibinfo {year}
  {2002})\BibitemShut {NoStop}%
\bibitem [{\citenamefont {Ngan}\ and\ \citenamefont {Shepherd}(1999)}]{ngan2}%
  \BibitemOpen
  \bibfield  {author} {\bibinfo {author} {\bibfnamefont {K.}~\bibnamefont
  {Ngan}}\ and\ \bibinfo {author} {\bibfnamefont {T.}~\bibnamefont
  {Shepherd}},\ }\href@noop {} {\bibfield  {journal} {\bibinfo  {journal}
  {Journal of the Atmospheric Sciences}\ }\textbf {\bibinfo {volume} {56}},\
  \bibinfo {pages} {4153} (\bibinfo {year} {1999})}\BibitemShut {NoStop}%
\bibitem [{\citenamefont {Waugh}\ and\ \citenamefont
  {Abraham}(2008)}]{Waugh2008}%
  \BibitemOpen
  \bibfield  {author} {\bibinfo {author} {\bibfnamefont {D.~W.}\ \bibnamefont
  {Waugh}}\ and\ \bibinfo {author} {\bibfnamefont {E.~R.}\ \bibnamefont
  {Abraham}},\ }\href@noop {} {\bibfield  {journal} {\bibinfo  {journal}
  {Geophysical Research Letters}\ }\textbf {\bibinfo {volume} {35}},\ \bibinfo
  {pages} {L20605} (\bibinfo {year} {2008})}\BibitemShut {NoStop}%
\bibitem [{\citenamefont {Michels}()}]{dislin}%
  \BibitemOpen
  \bibfield  {author} {\bibinfo {author} {\bibfnamefont {H.}~\bibnamefont
  {Michels}},\ }\href@noop {} {\enquote {\bibinfo {title} {Dislin home page},}\
  }\bibinfo {howpublished} {\url{http://www.mps.mpg.de/dislin/}, accessed May
  2010}\BibitemShut {NoStop}%
\bibitem [{\citenamefont {Waugh}\ \emph {et~al.}(2006)\citenamefont {Waugh},
  \citenamefont {Abraham},\ and\ \citenamefont {Bowen}}]{Waugh2006}%
  \BibitemOpen
  \bibfield  {author} {\bibinfo {author} {\bibfnamefont {D.}~\bibnamefont
  {Waugh}}, \bibinfo {author} {\bibfnamefont {E.}~\bibnamefont {Abraham}}, \
  and\ \bibinfo {author} {\bibfnamefont {M.}~\bibnamefont {Bowen}},\
  }\href@noop {} {\bibfield  {journal} {\bibinfo  {journal} {Journal of
  Physical Oceanography}\ }\textbf {\bibinfo {volume} {36}},\ \bibinfo {pages}
  {526} (\bibinfo {year} {2006})}\BibitemShut {NoStop}%
\bibitem [{\citenamefont {Fereday}\ and\ \citenamefont
  {Haynes}(2004)}]{Fereday04}%
  \BibitemOpen
  \bibfield  {author} {\bibinfo {author} {\bibfnamefont {D.}~\bibnamefont
  {Fereday}}\ and\ \bibinfo {author} {\bibfnamefont {P.}~\bibnamefont
  {Haynes}},\ }\href@noop {} {\bibfield  {journal} {\bibinfo  {journal}
  {Physics of Fluids}\ }\textbf {\bibinfo {volume} {60}},\ \bibinfo {pages}
  {4359} (\bibinfo {year} {2004})}\BibitemShut {NoStop}%
\bibitem [{\citenamefont {Haynes}\ and\ \citenamefont
  {Vanneste}(2005)}]{Haynes05}%
  \BibitemOpen
  \bibfield  {author} {\bibinfo {author} {\bibfnamefont {P.}~\bibnamefont
  {Haynes}}\ and\ \bibinfo {author} {\bibfnamefont {J.}~\bibnamefont
  {Vanneste}},\ }\href@noop {} {\bibfield  {journal} {\bibinfo  {journal}
  {Physics of Fluids}\ }\textbf {\bibinfo {volume} {17}},\ \bibinfo {pages}
  {097103} (\bibinfo {year} {2005})}\BibitemShut {NoStop}%
\bibitem [{\citenamefont {Tsang}(2009)}]{tsang09}%
  \BibitemOpen
  \bibfield  {author} {\bibinfo {author} {\bibfnamefont {Y.}~\bibnamefont
  {Tsang}},\ }\href@noop {} {\bibfield  {journal} {\bibinfo  {journal}
  {Physical Review E}\ }\textbf {\bibinfo {volume} {80}},\ \bibinfo {pages}
  {026305} (\bibinfo {year} {2009})}\BibitemShut {NoStop}%
\end{thebibliography}%

\end{document}